\def\im{{\rm im}}
\def\ip#1#2{\langle #1|#2\rangle}
\def\bra#1{\langle #1|}
\def\ket#1{|#1\rangle }
\def\complex{\bf C}
\def\frac#1#2{{#1\over #2}}
\def\fract#1#2{\textstyle{#1\over #2}}

\def\thm#1{\advance\count31 by 1\bigskip\noindent{\bf #1 \the\count31.}}
\def\pf{\medskip\noindent{\it Proof: }}
\def\qed{\quad{\vrule height8pt width5pt depth1pt}}

\def\sec#1{\bigbreak\advance\count32 by 1 
\bigskip\noindent{\bf \the\count32. #1}\nobreak\smallskip\noindent}
\def\ref{\advance\count33 by 1 \item{\the\count33.}}

\count31=0
\count32=-1
\count33=0

\centerline{{\bf Non-commutative Twistor Space}}

\bigskip
\centerline{K.C.\ HANNABUSS}

\medskip
\centerline{\it Balliol College, Oxford, OX1 3BJ, England. 
e-mail: kch@ermine.ox.ac.uk}



\bigskip\noindent
{\bf Abstract:} 
{It is shown that deformations of twistor space 
compatible with the Moyal deformation of Minkowski space-time must take
the 
form recently suggested by Kapustin, Kuznetsov and Orlov, [4]}

\bigskip
\sec{Introduction}

In an important contribution to the burgeoning study of integrable systems 
in non-commutative geometry (see [3], [6], [8]),  Kapustin, Kuznetsov and 
Orlov have generalised the Penrose-Ward twistor transform to that setting,
[4].
The geometric parts of the construction translate fairly naturally using
the standard techniques of non-commutative geometry ([2],[5]), but the form
(or even existence) of the non-commutative twistor algebra is less obvious.
In [4] such an algebra is presented using braiding properties defined by
an appropriate $R$-matrix.
In this note we consider the extent to which this is determined by the 
deformation of the original space-time algebra using a dual quantum group 
description.
The ordinary twistor space and the six-dimensional conformal space in
which compactified Minkowski space-time is identified with the rays of a
cone are both flag manifolds generated by the conformal group action on 
highest weight vectors, so we consider non-commutative twistor space as a 
quantum flag manifold for the quantised function algebra of a deformation 
of the conformal group [1].
We shall also give a coordinate-free expression for the $R$-matrix.

\sec{The non-commutative space-time algebra}

We recall that the non-commutative algebra for (four-dimensional) space-time 
${\cal M}$ is the twisted group algebra defined by a symplectic form
$\theta$ on ${\cal M}$. 
Its Lie algebra is generated by elements $X(u)$ for $u\in {\cal M}$ which 
satisfy $[X(u),X(v)] = i\hbar\theta(u,v)I$.
where $I$ is a positive central element.
Using the non-degenerate symmetric bilinear form $G$ on ${\cal M}$ 
we may write $i\hbar\theta(u,v) = G(u,\Theta v)$ for a suitable
skew-symmetric operator $\Theta$,
(which may be regarded as an element of the Lie algebra of the orthogonal 
group of $G$), so that 
$$[X(u),X(v)] = G(u,\Theta v)I.$$
We can also define $H = \frac12G^{ij}X_iX_j$ (where $X(u) = X_ju^j$ in
terms of coordinates and using the summation convention), and calculate that
$$[H,X(u)] = X(\Theta u)I.$$
In order to generalise the conformal theory it is useful to introduce a 
central element $T$ such that $I = T^2$, and to define $D = T^{-1}H$, so 
that the commutation relations take the homogeneous form 
$$[X(u),X(v)] =  G(u,\Theta v)T^2, \qquad
[D,X(u)] = X(\Theta u)T,$$
defining a quadratic algebra.
The identity 
$2TD = 2H = G^{ij}X_iX_j$,
gives the non-commutative version of the quadric cone of the conformal 
theory.

In the commutative case one studies connections with self-dual curvature.
This curvature vanishes on isotropic planes and these define twistors 
(projective spinors for the conformal group) ${\cal T}$, [7].
(For an isotropic plane through the origin, we note that the Clifford
algebra of ${\cal M}$ acts on the space of spinors.
Writing $\gamma(u)$ for the action of $u\in {\cal M}$ we may consider the
\lq\lq vacuum\rq\rq\ vector $\Psi_\xi\in {\cal T}$ with the property that 
$\gamma(u)\Psi_\xi = 0$ for all $u$ in the plane defined by 
$\xi\in \wedge^2{\cal M}$, that is for $u\wedge\xi = 0$.
Since $\Psi$ is determined only up to multiples this gives a ray, and
defines a point of the projective space of ${\cal T}$.
Isotropic planes through other points can be obtained using the action of
the conformal group.)

The space ${\cal T}$ decomposes into odd and even spinors 
${\cal T}_+\oplus{\cal T}_-$.
Each of the spaces ${\cal T}_\pm$ is two-dimensional so that 
$\wedge^2{\cal T}_\pm$ is spanned by a single vector $\epsilon_\pm$.
(We normalise these so that $\|\epsilon_\pm\|^2 = 2$, so that when we
choose orthonormal bases $e_1$, $e_2$ for ${\cal T}_+$ and $e_3$, $e_4$ for 
${\cal T}_-$ we may take 
$\epsilon_+ = e_1\wedge e_2 = e_1\otimes e_2 - e_2\otimes e_1$ and 
$\epsilon_- = e_3\wedge e_4 = e_3\otimes e_4 - e_4\otimes e_3$.)

The six-dimensional space $\wedge^2{\cal T}$ has a natural symmetric
bilinear form $B$ defined by
$$B(\xi,\eta)\epsilon_+\wedge\epsilon_- = \xi\wedge\eta.$$
The action of the conformal group on $\wedge^2{\cal T}$ leaves invariant
the null cone $B(u,u) = 0$, and the rays of this cone form a compact complex 
version of space-time.
(The finite points of space time can be identified with the rays of 
the cone for which $B(\epsilon_-,u) \neq 0$, or equivalently ${\cal M}$
can be identified with ${\cal T}_+\wedge{\cal T}_-$.
For brevity we shall call $\wedge^2{\cal T}$ conformal space-time.)

By construction the orthogonal group of $G$ has its spin representation 
$\Gamma$ on ${\cal T}$, which leaves each of ${\cal T}_\pm$ invariant, and
so the forms $\epsilon_\pm$ are also invariant under the spin action.
We may choose the orthonormal bases $e_1$, $e_2$ for ${\cal T}_+$ and 
$e_3$, $e_4$ for ${\cal T}_-$ so that the matrix $\Gamma(\Theta)$ is
diagonal:
$$\fract12\hbar\left(\matrix{(b-a)\sigma_3 &0\cr 
0 &(a+b)\sigma_3\cr}\right) \qquad{\rm where}\qquad 
\sigma_3 = \left(\matrix{1 &0\cr 0&-1\cr}\right).$$
We now complete $\epsilon_\pm$ to a basis for $\wedge^2{\cal T}$ by
introducing 
$\epsilon_1+i\epsilon_4 = e_4\wedge e_1$,
$\epsilon_1-i\epsilon_4 = e_2\wedge e_3$,
$\epsilon_2+i\epsilon_3 = e_1\wedge e_3$, and 
$\epsilon_2-i\epsilon_3 = e_2\wedge e_4$.
With this notation $\{\epsilon_\mu\}$ forms a $G$-orthonormal basis of 
${\cal M}\sim \wedge^2{\cal T}$ when we identify $G$ with $-2B$.
The matrix of $\Theta$ with respect to this basis is given by
$$i\hbar\left(
\matrix{0 &0 &0 &a\cr  0 &0 &b &0\cr 0 &-b &0 &0\cr -a &0 &0
&0\cr}\right).$$

Unfortunately, the commutation relations of the non-commutative theory
destroy the full conformal symmetry.
For example, the transformations of the space ${\cal M}$, affecting only
the operators $X(u)$, must not only be orthogonal with respect to $G$ but also
be symplectic with respect to $\theta$, or equivalently commute with
$\Theta$.
In general this forces them to lie in a Cartan subgroup, though, for 
particular $\theta$, (when $a$ vanishes) they may lie in a larger group. 
We must therefore deform the conformal group, or more precisely we shall 
quantise its function algebra by deforming the $R$ matrix associated with
the commutation relations.
To make explicit the connection between commutation relations and an $R$ 
matrix, we consider for any vector space $U$ an operator 
$R:U\otimes U \to U\otimes U$ satisfying the Yang-Baxter equation, and
write $R= \Phi +R^\prime$ where $\Phi$ denotes the usual flip operator 
$x\otimes y \mapsto y\otimes x$ on $U\otimes U$, and $R^\prime$ is the 
deformation. 
The function algebra defined by $R$ is given by operators $A$ on $U$ which 
satisfy $R(A\otimes A) = (A\otimes A)R$.
The following is then a straightforward consequence of our definitions.

\thm{PROPOSITION}
The relations $R(A\otimes A) = (A\otimes A)R$  of the function algebra 
defined by $R= \Phi +R^\prime$ are equivalent to the commutation relations 
$$[\ip{e}{Au},\ip{f}{Av}] 
= \ip{f\otimes e}{(A^{(2)}R^\prime-R^\prime A^{(2)})(u\otimes v)},$$
where $A^{(2)} = A\otimes A$, $e$, $f \in U^*$, $u$, $v \in U$, and 
$\ip{e}{u}$ denotes the pairing between the dual spaces $U^*$ and $U$.

\pf
The relations $R(A\otimes A) = (A\otimes A)R$ can be rewritten as
$$\Phi(A\otimes A) - (A\otimes A)\Phi = 
(A\otimes A)R^\prime- R^\prime(A\otimes A).$$
Writing $A^{(2)} = A\otimes A$, and choosing 
$e$, $f \in U^*$, $u$, $v \in U$ this gives
$$\ip{f\otimes e}{(\Phi A^{(2)}-A^{(2)}\Phi)(u\otimes v)}
= \ip{f\otimes e}{(A^{(2)}R^\prime-R^\prime A^{(2)})(u\otimes v)}.$$
The left-hand side can be written as
$$\ip{e\otimes f}{A^{(2)}u\otimes v} - \ip{f\otimes e}{A^{(2)}v\otimes u}
= \ip{e}{Au}\ip{f}{Av} - \ip{f}{Av}\ip{e}{Au} = [\ip{e}{Au},\ip{f}{Av}],$$
so that we have the stated result.
\qed

\bigskip
Let us take $U=\wedge^2{\cal T}$ and write the vectors as 
$v=v_+\epsilon_+ + v_0 + v_-\epsilon_-$, with
$v_0\in {\cal T}_+\wedge{\cal T}_- \sim {\cal M}$.
To avoid confusion with the twistor examples we write $\widetilde{R}$ and 
$\widetilde{A}$ instead of $R$ and $A$.
We are  less interested in the whole function algebra than in the quantum
flag manifold generated by the highest weight vector $\epsilon_+^*$ of the 
dual basis of $U^*$, and  want to 
identify $\ip{\epsilon_+^*}{\widetilde{A}v}$ with $v_+D +X(v_0) + v_-T$.
In this notation the commutation relations reduce to
$$[u_+D +X(u_0) + u_-T,v_+D +X(v_0) + v_-T]
=  u_+TX(\Theta v_0) -v_+X(\Theta u_0)T +G(u_0,\Theta v_0)T^2.$$
One simple way of obtaining these is to consider the deformation given by
$\widetilde{R}^\prime$ 
$$\widetilde{R}^\prime(u\otimes v) =
{u_+\epsilon_-\otimes \Theta v_0} - {\Theta u_0\otimes v_+\epsilon_-}
+G(u_0,\Theta v_0){\epsilon_-\otimes \epsilon_-}.$$
Since this has no terms in $\epsilon_+$, we have 
$\bra{\epsilon_+^*\otimes\epsilon_+^*}\widetilde{R}^\prime = 0$
and one may readily check that $\widetilde{R}$ gives the correct 
commutation relation.
It is useful to note that
$$\widetilde{R}^\prime(u\wedge v) = {u_+\epsilon_-\otimes_S\Theta v_0} 
-{\Theta u_0\otimes_S v_+\epsilon_-}
+ G(u_0,\Theta v_0){\epsilon_-\otimes_S \epsilon_-},$$
and 
$$\widetilde{R}^\prime(u\otimes_S v) = {u_+\epsilon_-\wedge\Theta v_0} 
-{\Theta u_0\wedge v_+\epsilon_-},$$
so that $\widetilde{R}^\prime$ interchanges symmetric and antisymmetric 
tensors.
It is also easy to check that the square of $\widetilde{R}^\prime$
vanishes.
These are in accord with the following simple result.

\thm{PROPOSITION}
A family of deformations $R= \Phi +R^\prime$ satisfies $R^2 =1$,  
if and only if $R^\prime$ maps the symmetric tensors to antisymmetric
tensors and vice versa, and ${R^\prime}^2 = 0$.

\pf
This is a matter of comparing terms of first and second order in
$R^\prime$ in the identity
$$1= R^2 = \Phi^2 + \Phi R^\prime + R^\prime\Phi + {R^\prime}^2 
= 1 + \Phi R^\prime + R^\prime\Phi + {R^\prime}^2,$$ 
which gives $\Phi R^\prime + R^\prime\Phi = 0 = {R^\prime}^2$.
The first identity can be rewritten in the form
$$R^\prime(1\pm \Phi) = (1\mp \Phi)R^\prime,$$
which, recalling that $\frac12(1 + \Phi)$ is the projection onto the 
symmetric tensor product $\otimes_S^2U$ and 
$\frac12(1 - \Phi)$ projects onto the antisymmetric tensor product 
$\wedge^2U$, shows that $R^\prime$ maps
$\otimes_S^2U$ to $\wedge^2U$ and 
$\wedge^2U$  to $\otimes_S^2U$.
(For similar reasons $R^\prime$ also interchanges
$\otimes_{S,R}^2U = \frac12(1+R)(U\otimes U)$ and  
$\wedge_R^2U = \frac12(1-R)(U\otimes U)$.)
\qed

\sec{Yang-Baxter operators for the twistors}

We now want to find a Yang-Baxter operator $R$ for twistor space which is 
consistent with that on conformal space-time $\wedge^2{\cal T}$.
More precisely, a Yang-Baxter operator $R$ on $\otimes^2{\cal T}^*$ can be 
extended to a Yang-Baxter operator 
$\widetilde{R}:\otimes^4{\cal T} \to \otimes^4{\cal T}$, defined by
$$\widetilde{R} = R_{23}R_{12}R_{34}R_{23}.$$
This interchanges the first and second pairs of factors in the tensor
product,
allowing one to extend the Yang-Baxter operator $R$ for twistor space to a 
Yang-Baxter operator $\widetilde{R}$ for the conformal space-time 
$\wedge_R^2{\cal T}$, and, for consistency, this should agree with that
found above.
The primary task is to see whether any such $R$ exist, and to find them.

It is easy to see that $\widetilde{R}$ inherits various properties of $R$,
but in the Appendix we show that two crucial properties of $\widetilde{R}$
must be shared by $R$, a fact which substantially simplifies the task of 
finding it.

\thm{THEOREM}
If a family of Yang-Baxter operators $R$ which is a deformation of the
flip 
$\Phi$ satisfies $[R_{23}R_{12}R_{34}R_{23}]^2 = 1$, then $R^2 = 1$, and
if, in addition, $R_{23}R_{12}R_{34}R_{23}$ commutes with 
$\Lambda^{(4)} = \Lambda\otimes\Lambda\otimes\Lambda\otimes\Lambda$ for
invertible 
$\Lambda$, then $R$ commutes with $\Lambda^{(2)} = \Lambda\otimes\Lambda$.

\bigskip
We have already observed that for conformal space-time $\widetilde{R}^2 =
1$, and the invariance of the commutation relations under the infinitesimal 
rotation $\Theta$ means that $\widetilde{R}$ commutes with $\Theta$ and so 
with the invertible operator $\Lambda = \exp(s\Theta)$.
This suggests that the non-commutative twistor algebra is defined by 
an operator $R$ satisfying $R^2 =1$, and commuting with the spin action of 
$\Lambda\otimes\Lambda = \exp(s\Theta_1+s\Theta_2)$, where 
$\Theta_1 = \Gamma(\Theta)\otimes 1$, and $\Theta_2 = 1\otimes
\Gamma(\Theta)$.
(It is merely suggested because the space-time $\widetilde{R}$ is only 
defined on the subspace $\wedge_R^2{\cal T}^*\otimes\wedge_R^2{\cal T}^*$.
There is no problem provided that $\widetilde{R}$ has an extension to the
whole of $\otimes^4{\cal T}^*$ with the same involutory and intertwining
properties.
Our argument will start by assuming that there is such an extension, and
then justifying this once $R$ has been found.
Once established we can deduce that $R$ also commutes with the spin action 
$\Theta_1+\Theta_2$.)
In fact $\widetilde{R}$ enjoys an additional, less obvious symmetry
arising from its compatibility with the exact sequence: 
$ 0 \longrightarrow {\cal T}_- \longrightarrow {\cal T}
\longrightarrow {\cal T}_+ \longrightarrow 0$.

\thm{PROPOSITION}
Let $N$ be a nilpotent operator on ${\cal T}$ such that 
$\im(N) = \ker(N) = {\cal T}_-$ and let $K_s = 1+sN$.
Then $K_s^{(4)} = K_s\otimes K_s\otimes K_s\otimes K_s$ commutes with 
$\widetilde{R}$.

\pf
By definition $K_s$ acts as the identity on ${\cal T}_-$ and so 
$K_s^{(2)}\epsilon_- = \epsilon_-$.
By contrast $K_s^{(2)}\epsilon_+$ must have the general form 
$\epsilon_+ + sn + s^2\nu\epsilon_-$ for scalar 
$\nu$ and $n\in {\cal T}_+\wedge{\cal T_-}$.
For $u_0\in {\cal T}_+\wedge{\cal T_-}$ we must have
$K_s^{(2)}u_0 = u_0 +s\lambda\epsilon_-$ for a suitable scalar $\lambda$
which we shall now find.
We first notice that the determinant of $K_s$ is clearly 1, so that for
any $\xi$ and $\eta\in {\cal T}\wedge{\cal T}$ we have
$\xi\wedge\eta = K_s^{(2)}\xi\wedge K_s^{(2)}\eta$, and thence 
$B(\xi,\eta) = B(K_s^{(2)}\xi,K_s^{(2)}\eta)$.
In particular, taking $\xi = \epsilon_+$ and $\eta = u_0$ gives
$$0 = B(\epsilon_+,u_0) = 
B(\epsilon_+ + sn + s^2\nu\epsilon_-,u_0 +s\lambda\epsilon_-)
= s\lambda +sB(n,u_0),$$
so that $\lambda = -B(n,u_0) = \fract12G(n,u_0)$, and 
$K_s^{(2)}u_0 = u_0 +\fract12sG(n,u_0)\epsilon_-$.

We now calculate that
$$\eqalign{K_s^{(4)}\widetilde{R}^\prime(u\otimes v) 
&= K_s^{(4)}\left(u_+\epsilon_-\otimes\Theta v_0 - \Theta u_0\otimes 
v_+\epsilon_- +G(u_0,\Theta v_0)\epsilon_-\otimes\epsilon_-\right)\cr
&= u_+\epsilon_-\otimes(\Theta v_0 +\fract12sG(n,\Theta v_0)\epsilon_-
- (\Theta u_0 +\fract12sG(n,\Theta u_0)\epsilon_-)\otimes 
v_+\epsilon_- +G(u_0,\Theta v_0)\epsilon_-\otimes\epsilon_-.\cr}$$
Since $G(n,\Theta u_0) = - G(u_0,\Theta n)$ and, in particular, 
$G(n,\Theta n) = 0$, this reduces to 
$$u_+\epsilon_-\otimes\Theta v_0 - \Theta u_0\otimes v_+\epsilon_- 
+G(u_0 + \fract12su_+n ,\Theta(v_0
+\fract12sv_+n))\epsilon_-\otimes\epsilon_-
= \widetilde{R}^\prime(K_s^{(4)}(u\otimes v)),$$
and since $\widetilde{\Phi}$ automatically commutes with any operator 
$\Lambda^{(4)}$ this gives the stated result.
\qed

We also want $R$ to satisfy the Yang-Baxter equation, but 
there is another requirement which is much stronger, namely that the 
restriction of $\widetilde{R} = R_{23}R_{12}R_{34}R_{23}$ to 
$\wedge_R^2{\cal T}^*\otimes\wedge_R^2{\cal T}^*$ contains only first
order terms in $\Theta$ (that is, in $R^\prime$).
With the requirements that $R$ be an involution and have the same
symmetries as $\widetilde{R}$, this limits the possibilities considerably.

\thm{THEOREM}
For non-degenerate $\Theta$, any Yang-Baxter operator $R = \Phi+R^\prime$  
satisfying $R^2 =1$, commuting with $\Theta$ and giving only first order 
corrections to $\widetilde{R}$  on the conformal spacetime is 
in the two-parameter family
$$R^\prime = \alpha\Theta_1\ket{\epsilon_-}\bra{\epsilon_+^*} 
+ \beta\ket{\epsilon_-}\bra{\epsilon_+^*}\Theta_1,$$
for some scalars $\alpha$, $\beta\in \complex$.
These operators automatically satisfy the Yang-Baxter equation.

\pf
Since $R^\prime$ commutes with the action of $\Theta$ it must preserve 
eigenspaces for the action.
The formula for $\Gamma(\Theta)$ shows that generically (unless $a$, $b$
or 
$a\pm b$ vanishes) each symmetric tensor $e_j\otimes e_j$ is in a 
one-dimensional eigenspace of its own, and since $R^\prime$ maps symmetric 
tensors to antisymmetric tensors we must have 
$e_j\otimes e_j\in\ker(R^\prime)$.
However, $\ker(R^\prime)$ must be invariant for operators of the form 
$K_s^{(2)}$ and by suitable choices of $N$, $K_se_j\otimes K_se_j$ 
($j=1,2$) generate the whole of ${\cal T}\otimes_S{\cal T}_-$ which must 
therefore also lie in the kernel.

Generically, there is also is a four-dimensional null space for the action
of $\Theta$ spanned by $e_1\otimes_Se_2$, $\epsilon_+ = e_1\wedge e_2$, 
$e_3\otimes_Se_4$, and $\epsilon_- = e_3\wedge e_4$, and the other
eigenspaces are two-dimensional spanned by tensors of the form 
$e_j\otimes_Se_k$ and $e_j\wedge e_k$ for various distinct $\{j,k\}
\notin \{\{1,2\},\{3,4\}\}$.
We have just seen that the symmetric tensors are in the kernel of
$R^\prime$.
Since $R^\prime$ maps antisymmetric to symmetric tensors each $e_j\wedge
e_k$ must map to a multiple of $e_j\otimes_Se_k$.
In fact, this multiple must also vanish otherwise 
overlapping terms such as 
$R_{12}^\prime R_{23}^\prime(e_i\otimes e_j\otimes e_k)$, give quadratic 
contributions to $\widetilde{R}$, so we may assume that 
$R^\prime(e^*_j\otimes_Se^*_k)=0$ and $R^\prime(e^*_j\wedge e^*_k)=0$ for
the given range of $j$ and $k$.

The only interesting deformations therefore occur in the four-dimensional
null space.
Here it is useful to note that 
$\Theta_1\epsilon_+ = \frac12\hbar(b-a)e_1\otimes_Se_2$,
and $\Theta_1\epsilon_- = \frac12\hbar(b+a)e_3\otimes_Se_4$.
(Since $\epsilon_\pm$ and their duals are invariant under the appropriate 
actions of $\Theta_1 +\Theta_2$, 
these could equally have been written in terms of $\Theta_2$.)
We already know that $R^\prime$ kills $e_3\otimes_Se_4$ and $e_3\wedge
e_4$, so, again discarding the deformations which mix the symmetric and
antisymmetric products of the same vectors and lead to quadratic 
contributions to $\widetilde{R}$, we are left with $R^\prime$ of the form 
$$R^\prime\epsilon_+= \alpha\Theta_1\epsilon_- \qquad
R^\prime\Theta_1\epsilon_+ = \alpha^\prime\epsilon_-,$$
or, equivalently, $R^\prime =
\alpha\Theta_1\ket{\epsilon_-}\bra{\epsilon_+^*} + 
\beta\ket{\epsilon_-}\bra{\epsilon_+^*}\Theta_1$.
The condition that ${R^\prime}^2 = 0$ holds automatically as do the 
Yang-Baxter equations.
\qed

\bigskip
We shall next show that with an appropriate choice of constants $\alpha$
and $\beta$ this is compatible with $\widetilde{R}$.
Since by construction $R_{23}R_{12}R_{34}R_{23}$ is an involution
commuting with the same operators as $\widetilde{R}$ this verifies that
there is indeed an extension of $\widetilde{R}$ to $\otimes^4{\cal T}^*$, 
so justifying our working assumption.

\thm{THEOREM}
The only Yang-Baxter operator for ${\cal T}$ compatible with the known 
Yang-Baxter operator $\widetilde{R}$ on the non-commutative conformal 
space-time is 
$$R = \Phi+ 2[\Theta_1,\ket{\epsilon_+^*}\bra{\epsilon_-}]_+.$$

\pf
We have noted that there are only first order corrections so the Yang-
Baxter operator takes the form
$\widetilde{R} = \widetilde{\Phi} + \widetilde{R}^\prime$, where 
$\widetilde{\Phi}$ is the flip and 
$$\widetilde{R}^\prime = \Phi_{23}\Phi_{12}\Phi_{34}{R}^\prime_{23} +
\Phi_{23}\Phi_{12}{R}^\prime_{34}\Phi_{23} +
\Phi_{23}{R}^\prime_{12}\Phi_{34}\Phi_{23} +
{R}^\prime_{23}\Phi_{12}\Phi_{34}\Phi_{23}.$$
Since, for example $\Phi_{23}R_{24} = R^\prime_{34}\Phi_{23}$, this can be 
rewritten as
$$\widetilde{R}^\prime = {R}^\prime_{14}\Phi_{23}\Phi_{12}\Phi_{34} +
{R}^\prime_{24}\Phi_{13} + {R}^\prime_{13}\Phi_{24} +
{R}^\prime_{23}\Phi_{12}\Phi_{34}\Phi_{23}.$$
Applying this to $(u\wedge x)\otimes (v\wedge y)$ we get
$$\eqalign{&{R}^\prime_{14}
(x\otimes y\otimes u\otimes v-u\otimes y\otimes x\otimes v
-x\otimes v\otimes u\otimes y+u\otimes v\otimes x\otimes y)\cr 
&+ {R}^\prime_{24}(v\otimes x\otimes u\otimes y-v\otimes u\otimes 
x\otimes y -y\otimes x\otimes u\otimes v+y\otimes u\otimes x\otimes v)\cr 
& + {R}^\prime_{13}(u\otimes y\otimes v\otimes x-x\otimes y\otimes
u\otimes v
-u\otimes v\otimes y\otimes x+x\otimes v\otimes y\otimes u)\cr 
& + {R}^\prime_{23}(v\otimes u\otimes y\otimes x-v\otimes x\otimes
y\otimes u
-y\otimes u\otimes v\otimes x+y\otimes x\otimes u\otimes v).\cr}$$
If we take $u$, $v\in {\cal T}_+$ and $x$, $y\in {\cal T}_-$ and use the
form of $R^\prime$ most of these terms disappear, leaving
$$-{R}^\prime_{14}u\otimes y\otimes x\otimes v +
{R}^\prime_{24}y\otimes u\otimes x\otimes v
+ {R}^\prime_{13}u\otimes y\otimes v\otimes x
-{R}^\prime_{23}y\otimes u\otimes v\otimes x.$$
It will be sufficient to consider the case of $u=e_1$ and $v=e_2$, where
$$\eqalign{R(e_1\otimes e_2) &= \fract12\hbar(a+b)\alpha e_3\otimes_S e_4
+\fract12\hbar(b-a)\beta e_3\wedge e_4\cr
&= \fract12\hbar(\alpha(a+b)+\beta(b-a))e_3\otimes e_4
+\fract12\hbar(\alpha(a+b)-\beta(b-a))e_4\otimes e_3.\cr}$$
Writing $\sigma = \fract12\hbar(\alpha(a+b)+\beta(b-a))$
and $\delta = \fract12\hbar(\alpha(a+b)-\beta(b-a))$,
we reduce  $\widetilde{R}^\prime(e_1\wedge x\otimes e_2\wedge y)$ to 
$$\eqalign{\sigma&(-e_3\otimes y\otimes x\otimes e_4 +
y\otimes e_3\otimes x\otimes e_4
+ e_3\otimes y\otimes e_4\otimes x -y\otimes e_3\otimes e_4\otimes x)\cr
&+\delta(-e_4\otimes y\otimes x\otimes e_3 +
y\otimes e_4\otimes x\otimes e_3
+ e_4\otimes y\otimes e_3\otimes x -y\otimes e_4\otimes e_3\otimes x)\cr
& =\sigma y\wedge e_3\otimes x\wedge e_4
+\delta y\wedge e_4\otimes x\wedge e_3.\cr}$$
This is non-vanishing if $x=e_3$ and $y=e_4$, or vice versa.
The first possibility gives
$$\widetilde{R}^\prime(e_1\wedge x\otimes e_2\wedge y) 
=\sigma e_4\wedge e_3\otimes e_3\wedge e_4 = -\sigma\epsilon_-
\otimes\epsilon_-,$$
whilst the second gives
$$\widetilde{R}^\prime(e_1\wedge e_4\otimes e_2\wedge e_3)
= -\delta \epsilon_-\otimes\epsilon_-.$$
On the other hand $e_1\wedge e_3 = \epsilon_2+i\epsilon_3$ and 
$e_2\wedge e_4 = \epsilon_2-i\epsilon_3$ 
gives
$$\widetilde{R}^\prime(e_1\wedge e_3 \otimes e_2\wedge e_4) =
G(e_1\wedge e_3, \Theta(e_2\wedge e_4))\epsilon_-\otimes\epsilon_-
= -2\hbar b\epsilon_-\otimes\epsilon_-.$$
This gives $\fract12(\alpha(a+b)+\beta(b-a)) = 2b$, so that we take
$\alpha = \beta = 2$.
This is also the choice appropriate to the other choice of $x$ and $y$ and 
leads to the stated solution, which expressed in terms of coordinates
agrees with that of [4].
We can also check that this form of $R^\prime$ works on tensors of the
form $\epsilon_+\otimes u_0$.
However, at this point we must mention a subtlety which has so far been 
suppressed, that we should really be working not with $\wedge^2{\cal T}$
but with $\wedge_R^2{\cal T} = (1-R)\otimes^2({\cal T})$.
Fortunately, the correction $R^\prime$ has a large kernel, and the only 
difference is for multiples of $\epsilon_+$, which should be replaced by
$\frac12(1-R)\epsilon_+ = \epsilon_+ -\Theta_1\epsilon_-$.  
Fortunately this does not cause any serious complications.
\qed

\sec{Commutation relations for twistors}

We can immediately combine Theorem 6 with Proposition 1 to 
give an explicit form for the twistor commutation relations.

\thm{COROLLARY}
The commutation relations for non-commutative twistor space can be written
as
$$\eqalign{\fract12[\ip{e}{Au},\ip{f}{Av}] 
&= \ip{f\otimes e}{A^{(2)}\Theta_1\epsilon_-}\ip{\epsilon_+^*}{u\otimes v}
+\ip{f\otimes e}{A^{(2)}\epsilon_-}\ip{\epsilon_+^*}{\Theta_1(u\otimes
v)}\cr
&\qquad-\ip{f\otimes e}{\Theta_1\epsilon_-}\ip{\epsilon_+^*}{Au\otimes Av}
-\ip{f\otimes e}{\epsilon_-}\ip{\epsilon_+^*}{\Theta_1(Au\otimes
Av)}.\cr}$$

\bigskip
Recalling that        
$\ker(R^\prime) \supseteq 
{\cal T}_-\otimes{\cal T} +{\cal T}\otimes{\cal T}_-$, 
we see that the first term on the right-hand side vanishes whenever $u$ or
$v$ lies in ${\cal T}_-$.
Similarly, the fact that $\im(R^\prime)\subseteq{\cal T}_-\otimes{\cal
T}_-$  means that the second term on the right vanishes for $e$ or $f$ in 
${\cal T}^*_+$.
Taken together we see that $[\ip{e}{Au},\ip{f}{Av}]=0$
whenever $e$ or $f \in {\cal T}^*_+$ and $u$ or 
$v\in {\cal T}_-$, and, in particular, elements of the form $\ip{e}{Au}$
with $e\in {\cal T}^*_+$ and $u\in {\cal T}_-$ are central in the algebra.

More generally it is sufficient for our purposes to use only $e$ and $f$ 
in ${\cal T}^*_+$, and then  we derive the commutation relation 
$$\eqalign{\fract12[\ip{e}{Au},\ip{f}{Av}] 
&= \ip{f\otimes e}{A^{(2)}\Theta_1\epsilon_-}\ip{\epsilon_+^*}{u\otimes v}
+\ip{f\otimes e}{A^{(2)}\epsilon_-}\ip{\epsilon_+^*}{\Theta_1(u\otimes
v)}.\cr}$$
Introducing the symplectic forms 
$\omega_\pm(u,v) = \ip{\epsilon_\pm}{u\otimes v}$ on ${\cal T}_\pm$, and 
using the self-adjointness of $\Theta_1$, we may 
rewrite this as
$$\eqalign{\fract12[\ip{e}{Au},\ip{f}{Av}] 
&= \ip{f\otimes e}{A^{(2)}\Theta_1\epsilon_-}\omega_+(u,v)
+\ip{f\otimes e}{A^{(2)}\epsilon_-}\omega_+(\Gamma(\Theta)u,v)\cr
&= \ip{f\otimes e}{A^{(2)}[\omega_+(u,v)\Theta_1 +
\omega_+(\Gamma(\Theta)u,v)]
\epsilon_-}.\cr}$$
The most interesting case arises for $u=e_1$ and $v=e_2$ when
$\omega_+(u,v) = 1$ and $\omega_+(\Gamma(\Theta)u,v) = \frac12\hbar(b-a)$,
so that
$$\eqalign{[\ip{e}{Ae_1},\ip{f}{Ae_2}] 
&= \hbar\ip{f\otimes e}
{A^{(2)}((b+a)e_3\otimes_S e_4+(b-a)e_3\wedge e_4)}\cr
&= 2\hbar\ip{f\otimes e}{A^{(2)}(be_3\otimes e_4+ae_4\otimes e_3)}.\cr}$$
Setting $A_{jr} = \ip{e_j}{Ae_r}$ and taking $e=e_r$ and $f=e_s$, we
deduce that whenever $r$ and $s$ are 1 or 2, the commutation relation gives
$$[A_{r1},A_{s2}] = 2\hbar \left(bA_{s3}A_{r4} + aA_{s4}A_{r3}\right).$$
Setting $z_j = A_{1j}$, $w_j = A_{2j}$ and recalling that we already know
that $z_3$, $z_4$, $w_3$, and $w_4$ are central, we obtain the relations
$$[z_{1},z_{2}] = 
2\hbar(a+b)z_{3}z_{4}
\qquad
[z_{1},w_{2}] = 
2\hbar\left(az_{3}w_{4}+bz_4w_3\right)
\qquad
[w_{1},w_{2}] = 
2\hbar(a+b)w_{3}w_{4}.$$
The first of these is essentially the relation of [4] for the twistor 
algebra.
These provide an alternative way of constructing the Minkowski algebra by 
rewriting the matrix elements $\ip{\epsilon^*_+}{A^{(2)}\xi}$ for $\xi\in 
\wedge_R^2{\cal T}^*$ in terms of $z_{jk} = w_jz_k-w_kz_j$,
we get the formulae of [4].
For example, $T = \ip{\epsilon_+^*}{A^{(2)}\epsilon_-} = z_3w_4 - w_3z_4$
is central to the algebra.

\bigskip\noindent{\bf Appendix}

\medskip
\bigskip
This Appendix is devoted to a proof of Theorem 3.
This is done in stages through a chain of propositions.
We start by rewriting the conditions that $\widetilde{R}^2=1$, and 
that $\widetilde{R}$ commutes with 
$\Lambda^{(4)} = \Lambda\otimes\Lambda\otimes\Lambda\otimes\Lambda$.
The first condition is equivalent to 
$\widetilde{R}= \widetilde{R}^{-1} = [{R^{-1}}]\,\widetilde{\,}$, and the
second to 
$\widetilde{R} = \Lambda^{(4)}\widetilde{R}{\Lambda^{(4)}}^{-1}
= [{\Lambda^{(2)}R{\Lambda^{(2)}}^{-1}}]\,\widetilde{\,}$.
We should like to see whether these two statements, together with the 
Yang-Baxter equations, are sufficient to give 
$R^{-1} = R = \Lambda^{(2)} R{\Lambda^{(2)}}^{-1}$.
 
To this end we ask whether two Yang-Baxter operators $P$ and $Q$ 
satisfying 
$$P_{23}P_{12}P_{34}P_{23} = Q_{23}Q_{12}Q_{34}Q_{23}$$
must necessarily be identical, that is $P = Q$.

\thm{LEMMA}
Yang-Baxter operators $P$ and $Q$ satisfy
$P_{23}P_{12}P_{34}P_{23} = Q_{23}Q_{12}Q_{34}Q_{23}$
if and only if there exist $Y$ and $Z$ such that
$$P_{23}P_{12} =Q_{23}Q_{12}Z_{23}, \qquad
Q_{23}Q_{12} = Z_{12}P_{23}P_{12},$$
$$P_{12}P_{23} = Q_{12}Q_{23}Y_{12},
\qquad Q_{12}Q_{23} = Y_{23}P_{12}P_{23}.$$

\pf
Our main tool is separation of variables, which we shall use repeatedly.
We start with our assumed identity, which can be rearranged as
$$Q_{12}^{-1}Q_{23}^{-1}P_{23}P_{12} =
Q_{34}Q_{23}P_{23}^{-1}P_{34}^{-1}.$$
Since the left hand side is independent of $4$ and the right is
independent of $1$, we see that each side must act as the identity on
those factors and so can be written in the form $Z_{23}$.
Rearranging this gives
$$P_{23}P_{12} =Q_{23}Q_{12}Z_{23}, \qquad
Q_{34}Q_{23} = Z_{23}P_{34}P_{23},$$
and by dropping the indices by 1 the second equation becomes
$$Q_{23}Q_{12} = Z_{12}P_{23}P_{12}.$$
Similarly, since $P_{12}$ and $P_{34}$ commute (and similarly for $Q$) 
the assumed identity can be rearranged as
$Q_{34}^{-1}Q_{23}^{-1}P_{23}P_{34} = Q_{12}Q_{23}P_{23}^{-1}P_{12}^{-1}$,
and setting both sides equal to $Y_{23}$ gives (with an index shift)
$$P_{12}P_{23} = Q_{12}Q_{23}Y_{12},
\qquad Q_{12}Q_{23} = Y_{23}P_{12}P_{23}.$$
It is clearly possible to recover the original equation by eliminating
either $Y$ or $Z$.
\qed

\thm{LEMMA}
Yang-Baxter operators $P$ and $Q$ satisfy 
$P_{23}P_{12}P_{34}P_{23} = Q_{23}Q_{12}Q_{34}Q_{23}$
if and only if there exist $Y$, $Z$, $C$ and $D$ satisfying 
$$Q_{12}Z_{12} =C_2P_{12}, \qquad Q_{12}Y_{12}= C_1P_{12}.$$
$$Z_{12}P_{12} = Q_{12}D_1, \qquad Y_{12}P_{12}= Q_{12}D_2,$$
and 
$P_{12}P_{23}P_{12} = Q_{12}Q_{23}Q_{12}D_2 = C_2^{-1}Q_{12}Q_{23}Q_{12}$.

\pf
Using the Yang-Baxter equation for $Q$ and the second equation of each
pair in Lemma 8 we obtain 
$$Q_{12}Z_{12}P_{23}P_{12} = Q_{12}Q_{23}Q_{12}
= Q_{23}Q_{12}Q_{23}
= Q_{23}Y_{23}P_{12}P_{23}.$$
Now using the Yang-Baxter equation for $P$, we deduce that
$$Q_{12}Z_{12}P_{12}^{-1} = Q_{23}Y_{23}P_{23}^{-1},$$
and, since one side is the identity on the first factor and the other on
the third, we deduce that both sides are of the form $C_2$ for some operator
$C$.
Rearranging this gives
$$Q_{12}Z_{12} =C_2P_{12}, \qquad 
Q_{23}Y_{23}= C_2P_{23},$$
and by lowering the indices the second equation gives
$$Q_{12}Y_{12}= C_1P_{12}.$$
Similarly using the first equation of each pair
$$Q_{23}Q_{12}Z_{23}P_{23}=
P_{23}P_{12}P_{23} =
P_{12}P_{23}P_{12}
= Q_{12}Q_{23}Y_{12}P_{12},$$
from which we see that
$$Q_{23}^{-1}Z_{23}P_{23}
= Q_{12}^{-1}Y_{12}P_{12},$$
and both sides must be of the form $D_2$, giving
$$Z_{12}P_{12} = Q_{12}D_1, \qquad Y_{12}P_{12}= Q_{12}D_2.$$
Substituting these expressions into the previous identities we obtain
$P_{12}P_{23}P_{12}
= Q_{12}Q_{23}Q_{12}D_2$,
and similarly
$P_{12}P_{23}P_{12} = C_2^{-1}Q_{12}Q_{23}Q_{12}$.
The argument easily reverses.
\qed

\bigskip
The next two results cover the cases of real interest to us.
For these we limit ourselves to Yang-Baxter operators $P$ and $Q$ which
are deformations of the standard flip operators, from which one readily 
checks that $Y$, $Z$, $C$ and $D$ are all deformations of the identity 
operator.
The main observation which we need is that each deformation of the
identity operator, such as $C$, has a unique square root which is also a 
deformation of the identity. It is given by the binomial expansion of 
$\sqrt{1+x}$ with $C-1$ substituted for $x$.
(In fact, since ${\cal T}$ is finite-dimensional, this can be rewritten as 
a polynomial in $C$).

\thm{LEMMA}
If Yang-Baxter operators $P$ and $Q$ satisfy
$P_{23}P_{12}P_{34}P_{23} = Q_{23}Q_{12}Q_{34}Q_{23}$
and $Q = P^{-1}$, then $Q = P$, that is $P^2 = 1$.

\pf
When $Q=P^{-1}$ we immediately get the equations
$$(P_{12}P_{23}P_{12})^2 = D_2 = C_2^{-1},$$
so that $C$ and $D$ are inverses.
Now we also have
$$D_2 = (P_{12}P_{23}P_{12})^2 =
(P_{12}P_{23}P_{12})(P_{23}P_{12}P_{23}) = 
(P_{12}P_{23})^3,$$
and similarly
$$D_2 = (P_{23}P_{12})^3.$$
Consequently
$$P_{12}D_2 = P_{12}(P_{23}P_{12})^3 
= (P_{12}P_{23})^3P_{12} = D_2P_{12},$$
so that $D_2$, and therefore $C_2$, commute with $P_{12}$.
They similarly commute with $P_{23}$, which, by lowering the indices,
means that $D_1$ and $C_1$ commute with $P_{12}$.
(Since $C_j$ clearly commutes with $P_{12}$ for $j>2$ this shows that all
the $C_j$ and $P_{rs}$ commute.)

The remaining equations of Lemma 9 now give
$$Q_{12}^2D_1 = Q_{12}Z_{12}P_{12} =C_2P_{12}^2,$$
and
$$Q_{12}^2D_2 = 
Q_{12}Y_{12}P_{12}= C_1P_{12}^2.$$
Given that $C$ and $D$ are inverse and commute with $P_{12}$, these reduce
to $P_{12}^4 = C_1C_2$, or $P^4 = C\otimes C$.
As noted above $C$ has a unique square root $K$ deformed from the
identity, and given that the undeformed value of $P^2$ is 1 we take
square roots to see that $P^2 = K\otimes K$.

Since $K$ is given by a polynomial in $C$ it enjoys the same commutation 
properties as $C$, and in particular $K_j$ commutes with each of the
$P_{rs}$.
This means that
$$\eqalign{K_2^{-2} &= C_2^{-1}\cr 
& = (P_{12}P_{23}P_{12})^2\cr
& = P_{12}P_{23}P_{12}^2P_{23}P_{12}\cr
& = P_{12}P_{23}K_1K_2P_{23}P_{12}\cr
& = K_1K_2P_{12}P_{23}^2P_{12}\cr
& = K_1K_2^2K_3P_{12}^2\cr
& = K_1^2K_2^3K_3.\cr}$$
Separating the indices we see that $K = 1$, and so $P^2 = K\otimes K = 1$.
\qed

\bigskip
We now look at the second case where 
$Q_{12} = \Lambda^{(2)}P_{12}{\Lambda^{(2)}}^{-1}$.
and assume that $P^2 = 1 = Q^2$.

\thm{LEMMA}
If Yang-Baxter operators $P$ and $Q$ satisfy
$P_{23}P_{12}P_{34}P_{23} = Q_{23}Q_{12}Q_{34}Q_{23}$
and $P^2 = 1 = Q^2$ then $P=Q$.

\pf
Since $P$ and $Q$ are involutions our earlier identity 
$C_2P_{12}^2 = Q_{12}^2D_1$
reduces to $C_2 = D_1$, so that $C$ and $D$ must both be multiples of 1, 
indeed the same multiple, $\lambda 1$.
The equation
$$Q_{12}Q_{23}Q_{12}D_2
= C_2^{-1}Q_{12}Q_{23}Q_{12}.$$
then shows that $\lambda^2 =1$, and using the undeformed value of $C =1$
we deduce that $C = 1 = D$.
This means that 
$$Q_{12}Z_{12} = P_{12}, \qquad Q_{12}Y_{12}= P_{12}.$$
$$Z_{12}P_{12} = Q_{12}, \qquad Y_{12}P_{12}= Q_{12},$$
and 
$$P_{12}P_{23}P_{12}
= Q_{12}Q_{23}Q_{12}.$$
The first sets of equations are now consistent and can be regarded simply
as definitions of $Y$ and $Z$.
Using them in the second set we have
$$Y_{23}P_{23} = Q_{23} 
= Q_{12}^{-1}P_{12}P_{23}P_{12}Q_{12}^{-1} = Y_{12}P_{23}Y_{12}^{-1}.$$
We now write  $P_{jk} = \widehat{P}_{jk}\Phi_{jk}$,
where $\Phi_{jk}$ is the flip which is the undeformed value of $P$.
Then
$$Y_{23}\widehat{P}_{23}\Phi_{23}Y_{12}
= Y_{12}\widehat{P}_{23}\Phi_{23},$$
and since
$\Phi_{23}Y_{12}= Y_{13}\Phi_{23}$, this gives
$$Y_{23}\widehat{P}_{23}Y_{13}= Y_{12}\widehat{P}_{23}$$
Similarly we obtain
$$Y_{12}\widehat{P}_{12}Y_{13}= Y_{23}\widehat{P}_{12},$$
and from these we deduce that
$$(\widehat{P}_{23}Y_{13}^{-1}\widehat{P}_{23}^{-1})
(\widehat{P}_{12}Y_{13}^{-1}\widehat{P}_{12}^{-1})
= (Y_{23}^{-1}Y_{12})(Y_{12}^{-1}Y_{23}) = 1,$$
and thence that
$$Y_{13}\widehat{P}_{23}^{-1}\widehat{P}_{12}Y_{13}= 
\widehat{P}_{23}^{-1}\widehat{P}_{12}.$$

Setting $V= \widehat{P}_{23}^{-1}\widehat{P}_{12}$, this can be rewritten 
as $Y_{13}VY_{13}=V$, which leads to 
$$Y_{13}V^2= Y_{13}VY_{13}VY_{13} = 
V^2Y_{13},$$
showing that $Y_{13}$ and $V^2$ commute.
Our standard square root argument shows that $Y_{13}$ commutes 
with $V$, which in turn gives the identity $Y_{13}^2 = 1$, and so 
$Y_{13} = 1$,  and $Q = P$.
In the case in hand this shows that $R$ commutes with $\Lambda$ and so
with $\Theta$. 
\qed

\bigskip\noindent
{\bf References}

\medskip
\ref
V.\ Chari \& A.\ Pressley {\it A guide to Quantum Groups}, Cambridge
University 
Press, 1994.

\ref
A.\ Connes, {\it Non-commutative geometry}, Academic Press, 1994.

\ref
A.\ Connes , M.R.\ Douglas and A.S.\ Schwarz, 
{\it Non-commutative geometry and matrix theory}, J. High Energy Phys.  
{\bf 2}, Paper 3 (1998), hep-th/9711162.

\ref
A.\ Kapustin, A.\ Kuznetsov, \& D.\ Orlov, 
{\it Non-commutative instantons and twistor transform}, Comm. math. Phys. 
{\bf 221} (2001) 385-432,  hep-th/ 0002193.

\ref 
Yu.I.\ Manin, {\it Topics in non-commutative geometry}, {Princeton UP},
Princeton, 1991.

\ref
N.\ Nekrasov and A.S.\ Schwarz,
{\it Instantons on non-commutative ${\bf R}^4$ and (2,0) superconformal 
six-dimensional theory}, Comm. math. Phys. 
{\bf 198} (1998) 689-703, hep-th/9802068.

\ref
R.\ Penrose and W.\ Rindler,
{\it Spinors in space-time, vol 2}, Cambridge University Press, 1986.

\ref
N.\ Seiberg and E.\ Witten,
{\it String theory and non-commutative geometry}, J. High Energy Phys. 
{\bf 3}, Paper 32 (1999), hep-th/9908142.

\bye